\documentclass[twocolumn,aps,prl,amssymb,amsmath,floatfix,showpacs]{revtex4}
\usepackage{graphicx,subfigure}
\usepackage{bm}
\usepackage{verbatim}
\usepackage{amsmath}
\usepackage{amssymb}
\usepackage[T1]{fontenc}
\usepackage{ae,aecompl}

\newcommand{\rv}{{\bf r}}
\newcommand{\rh}{{\hat r}}
\newcommand{\vh}{{\hat v}}

\newcommand{\kv}{{\bf k}}

\newcommand{\vv}{{\bf v}}

\newcommand{\zh}{{\hat{\bf z}}}

\newcommand{\grad}{{\bm{\nabla}}}

\def\rf#1{(\ref{#1})}

\begin{document}

\title{Anomalous energetics and dynamics of moving vortices}
\author{Leo Radzihovsky} 
\affiliation{Department of Physics,
  University of Colorado, Boulder, CO 80309}
\date{\today}

\begin{abstract}
  Motivated by the general problem of {\em moving} topological defects
  in an otherwise ordered state and specifically, by the anomalous
  dynamics observed in vortex-antivortex annihilation and coarsening
  experiments in freely-suspended smectic-C films\cite{ClarkExp}, I
  study the deformation, energetics and dynamics of moving vortices in
  an overdamped xy-model and show that their properties are
  significantly and qualitatively modified by the motion.
\end{abstract}

\maketitle

{\em Introduction:} Topological defects play a central role in phase
transitions, relaxation of generalized strain in ordered states (e.g.,
current dissipation in a superfluid, strain relaxation in a
crystalline solid, etc.)\cite{LRT86}, coarsening dynamics after a
quench into an ordered state\cite{YurkeHusePRE93}, and appear in a
broad range of physical realizations from superfluids and liquid
crystals\cite{CLbook} to early universe
baryogenesis\cite{ChuangScience}.

Many physical systems involve topological defects moving
(stochastically or deterministically) through an otherwise ordered
medium. Although it is usually tacitly assumed that defect's
properties (texture structure, interaction, dynamics, etc) are not
modified by its motion, with the center $\rv_0$ simply boosted
$\rv_0\rightarrow \rv_0(t)$ by the motion, there is no a priori reason
for this to be the case. Instead, not unlike a relativistic charged
particle, a moving defect is defined and governed by the dynamics of
the associated vector field, requiring a nontrivial analysis that is
the subject of this Letter.

Stimulated by this general question, and by the anomalous
vortex-antivortex annihilation and coarsening
dynamics\cite{YurkeHusePRE93} observed in freely-suspended smectic-C
films experiments\cite{ClarkExp,PleinerPRA88,MuznyClarkPRL92}, I
explored the nature of moving vortices in an overdamped
two-dimensional (2D) xy-model, applicable to a broad range of soft
matter systems. In this Letter I report the results of these studies,
that, with some modifications may also extend to vortices in a
nonzero-temperature superfluid and superconductor in the presence of a
background supercurrent or dislocations in a strained crystal.

\begin{figure}[tbp]
\includegraphics[height=1.in,width=2.8in]{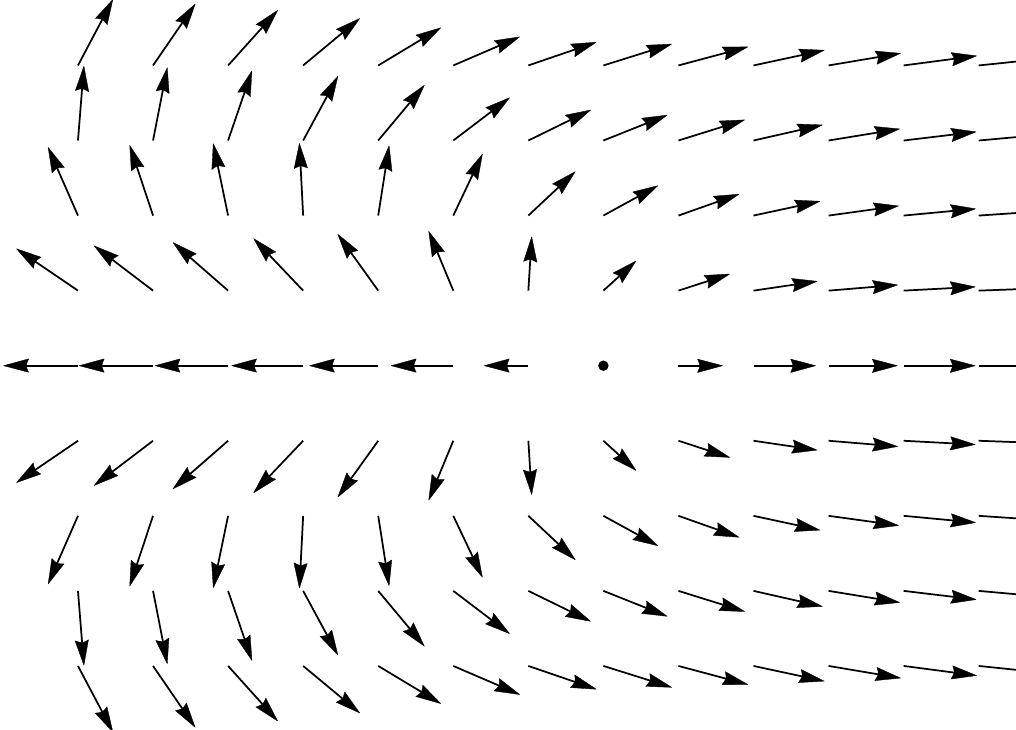}
\includegraphics[height=1.in,width=2.8in]{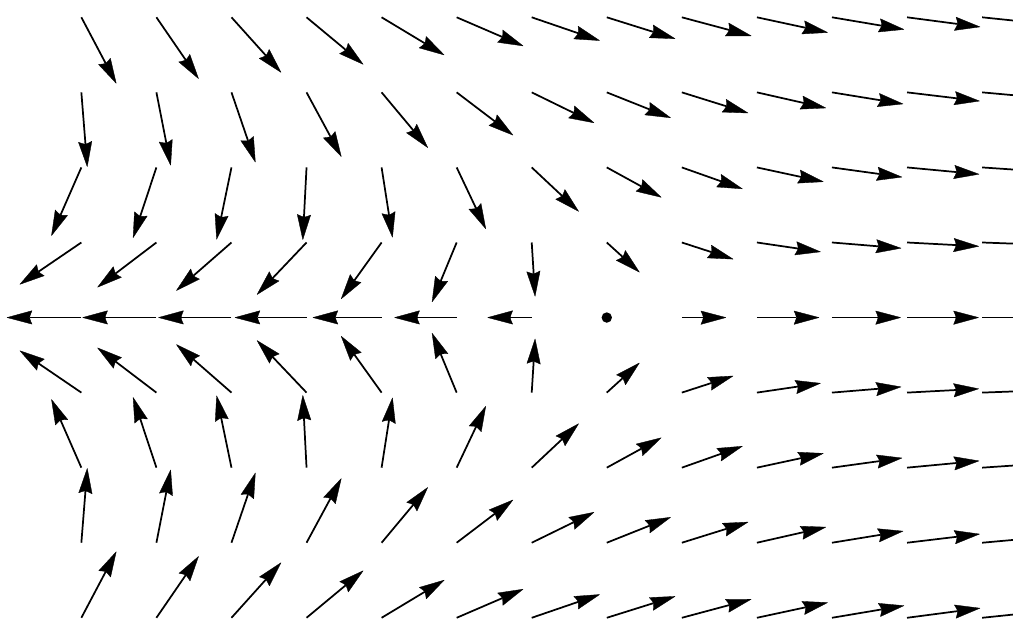}
\caption{A vector field corresponding to a phase $\theta(\rv)$ for a
  vortex-antivortex pair co-moving to the right with velocity $\vv$.}
\label{fig:vortexantivortexComoving}
\end{figure}

\begin{figure}[tbp]
\includegraphics[height=1.5in,width=3in]{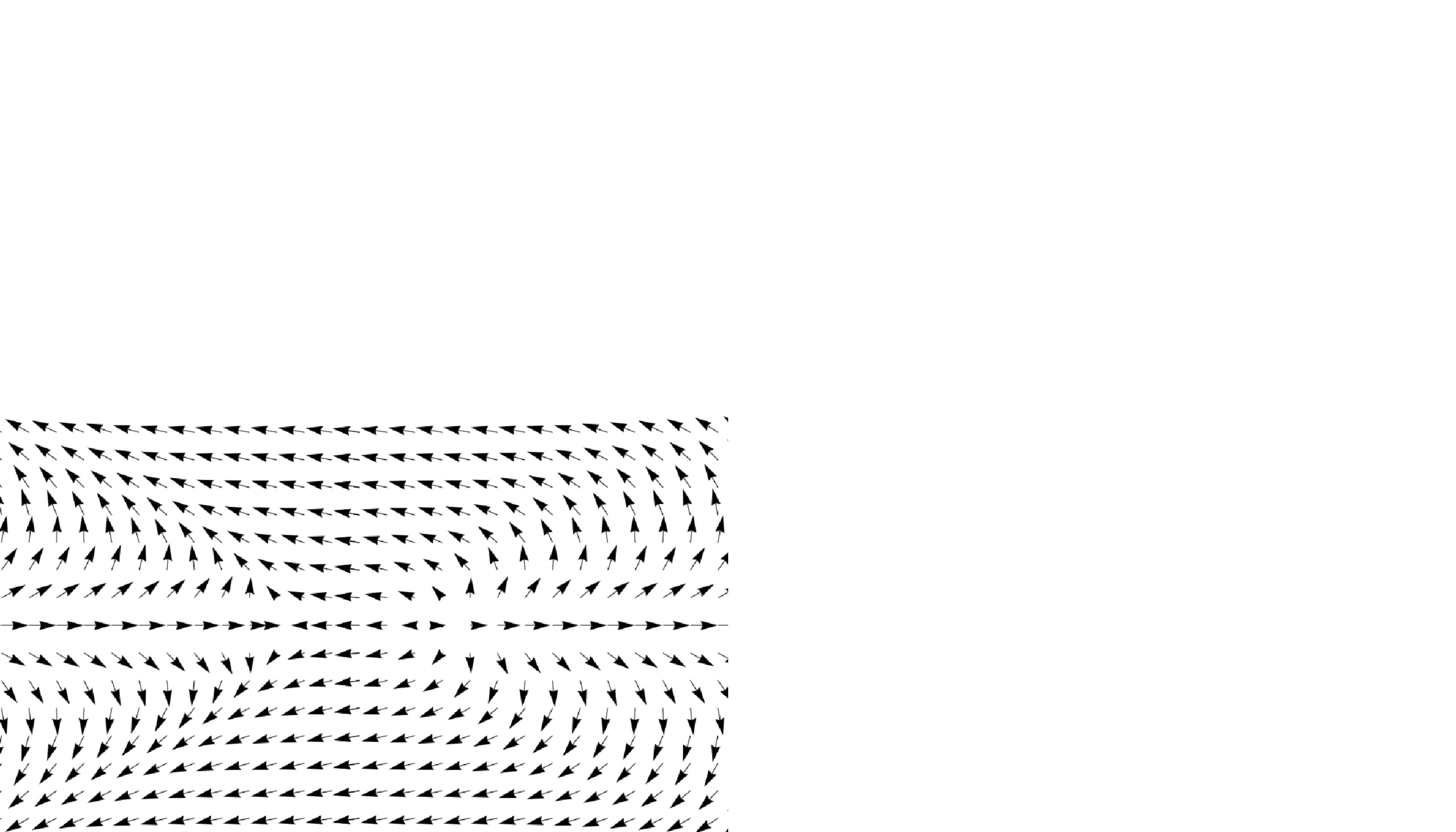}
\caption{A vortex-antivortex counterpropagating pair relevant to the
  annihilation problem. The motion-induced comet-like tails (that lead
  to a linearly diverging vortex elastic energy) and a suppression of
  the deformation between vortices (that leads to a weakened
  interaction) are clearly visible.}
\label{fig:vortexantivortexAnnihilate}
\end{figure}

{\em Results:} Before turning to the analysis, I summarize the results
of this study. I find that a vortex imposed to move with a constant
velocity $v$ in an ordered medium of stiffness $K$ and damping
$\gamma$, beyond a length scale 
\begin{eqnarray}
\xi_v = \frac{K}{\gamma v}\equiv D/v\equiv k_v^{-1}
\end{eqnarray}
exhibits a nontrivial longitudinal distortion of its standard, purely
transverse form\cite{CLbook}, latter retained on length scale below
$\xi_v$.  In the {\em steady state} the resulting deformed vortex
exhibits a parabolic comet-like tail, extending across the system to
which most of the $2\pi$ phase winding is confined (see
Figs.\ref{fig:vortexantivortexComoving},\ref{fig:vortexantivortexAnnihilate}).
While a motion-induced distortion is not surprising, the {\em
  qualitative} long-scale nature of its consequences (see below)
indeed is.
For a {\em transient state} at time $t$ after a vortex begins to move,
the steady-state distortion only extends out to a time-dependent
anisotropic ``horizon'' $v t \times\sqrt{K t/\gamma}$, beyond which
the purely transverse vortex field is nearly undistorted by the
motion. This is analogous to the Lienard-Wiechert potential of a
moving point charge\cite{Jackson}.
All other predictions follow from this result. Specifically, the
vortex steady-state mobility
\begin{eqnarray}
\mu_v\approx\frac{1}{\pi\gamma\ln(2\xi_v/a)}\sim 1/|\ln v|
\label{mu_result}
\end{eqnarray}
vanishes logarithmically with vanishing velocity, $\xi_v$ cutting off
the $\ln L/a$ divergence of a stationary vortex drag coefficient, a
result that was previously found via scaling and numerical analysis in
earlier
studies\cite{JProst83,LRT86,RyskinKremenetskyPRL91,YurkeHusePRE93}
($a$ is the vortex core size). Thus, a 2D vortex exhibits a breakdown
of a linear response to an external force $f$, with a truly nonlinear
velocity-force characteristics $v(f)\sim f/|\ln f|$.

\begin{figure}[tbp]
\includegraphics[height=1.5in,width=3in]{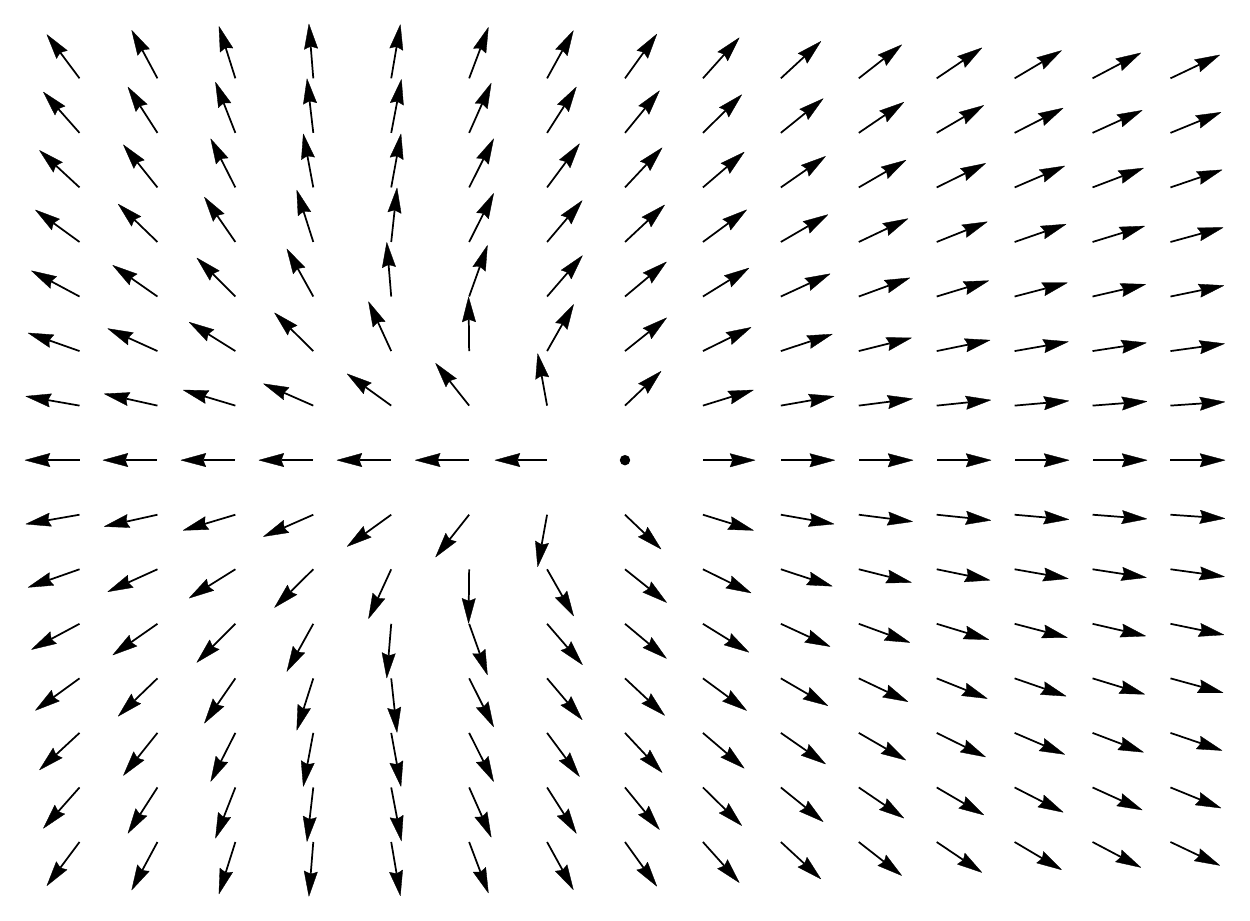}
\caption{A transient vector field of a vortex moving for time $t$,
  exhibiting a steady-state distortion out to an elliptical
  ``horizon'' $v t \times\sqrt{K t/\gamma}$, and purely transverse
  vortex field beyond.}
\label{fig:vortex_transient}
\end{figure}

The ``comet tail'' texture of a moving vortex leads to an elastic
energy that diverges linearly with system size
\begin{eqnarray}
E_v&\approx&\pi K\left(L/\xi_v + \ln\xi_v/a\right),
\label{Evortex}
\end{eqnarray}
with the usual logarithm cut off by the length $\xi_v\sim 1/v$, that
diverges with a vanishing velocity. The interaction between two moving
vortices strongly depends on their velocities and orientation relative
to the separation vector, $\rv$. With the eye on the problems of a
vortex-antivortex annihilation and nucleation by an imposed strain, I
find the interaction potential $U^\parallel_{\vv,-\vv}(r)$, for a
vortex and antivortex moving toward each other, $\vv\parallel\rv$
(Fig.\ref{fig:vortexantivortexAnnihilate}):
\begin{eqnarray}
U^\parallel_{\vv,-\vv}(r)&=&2\pi K\left[c-\sinh^{-1}(\xi_v/r)\right],\\
&\approx&2\pi K\left\{\begin{array}{ll}
c - \xi_v/r,&a\ll\xi_v\ll r,\nonumber\\
\ln r/a,&a\ll r\ll\xi_v,\\
\end{array}\right.
\label{Uparallel-vv}
\end{eqnarray}
and a potential $U^\perp_{\vv,\vv}(r)$ for a pair co-moving with velocity
$\vv\perp\rv$ (Fig.\ref{fig:vortexantivortexComoving}):
\begin{eqnarray}
  U^\perp_{\vv,\vv}(r)&=&2\pi K
  \left[c-\sinh^{-1}(\xi_v/r)+\sqrt{r^2/\xi_v^2+1}\right],\ \ \\
&\approx&2\pi K\left\{\begin{array}{ll}
c + r/\xi_v,&a\ll\xi_v\ll r,\nonumber\\
\ln r/a,&a\ll r\ll\xi_v,\\
\end{array}\right.
\label{Uperpvv}
\end{eqnarray}
with $c=\ln \xi_v/a$. Thus, in the annihilation configuration
(Fig.\ref{fig:vortexantivortexAnnihilate}), vortex attraction for
separation beyond $\xi_v$ is suppressed by the motion. Conversely and
even more dramatically, I predict that vortex pair motion in the
transverse configuration (Fig.\ref{fig:vortexantivortexComoving}) leads to a
{\em linear confinement} on long scales.

The above velocity-dependent vortex mobility and interaction
qualitatively modify the equation of motion for the vortex-antivortex
separation. This leads to a late-time slowed annihilation dynamics
that may be an important ingredient in the anomalies observed in the
experiments\cite{ClarkExp}.

{\em Analysis:} With the above motivation in mind, I now turn to the
analysis of moving vortices in a 2D overdamped xy-model 
\begin{eqnarray}
\gamma\partial_t\theta = K\nabla^2\theta,\ \ \mbox{\text with}\ \ \grad\times\grad\theta = 2\pi\delta(\rv-\rv_v(t))\zh,
\label{eom}
\end{eqnarray}
searching for a vortex solution $\theta(\rv,t)$, that for simplicity I
take to be moving at constant velocity defined by $\rv_v(t) = \vv t$.
Despite ignoring a number of ingredients\cite{neglectEffects}, I
expect it to be a core description of many systems where damping is
dominant.

To this end, I take the solution to be $\theta(\rv,t)=\theta_v(\rv-\vv
t) + \theta_s(\rv,t)$, where $\theta_v(\rv) =\varphi =\arctan(y/x)$ is
the azimuthal polar angle that is the standard purely transverse
solution of the static problem ($\gamma = 0$), that enforces a moving
unit of vorticity. The $\theta_s(\rv,t)$ part is a nonsingular,
single-valued function (with a purely longitudinal, curl-free
gradient) determined by the requirement that $\theta(\rv,t)$ satisfies
the equation of motion \rf{eom}. Thus $\theta_s(\rv,t)$ describes the
distortion of a moving vortex about the stationary form
$\theta(\rv)=\varphi$, with its spatial Fourier transform satisfying
\begin{eqnarray}
  \gamma\partial_t\theta_s(\kv) + K k^2\theta_s(\kv) &=&
\gamma\vv\cdot\grad\theta_v(\kv)e^{-i\kv\cdot\vv t}.
\label{eom_theta_s}
\end{eqnarray}
The exact solution is easily found either directly for
$\theta_s(\rv,t)$ or by first Galilean-transforming to the moving
vortex frame $\rv'=\rv-\vv t$, $\partial_t\rightarrow\partial_t +
\vv\cdot\nabla_{\rv'}$, where the distortion is
$\theta_s'(\rv',t)\equiv\theta_s(\rv'+\vv t,t)$.

For a vortex that has been moving forever the Fourier transform of the
steady state distortion (vanishing for $v=0$) is given by
\begin{eqnarray}
  \theta_s'(\kv) &=& \frac{\gamma\vv\cdot\grad\theta_v(\kv)}
  {K k^2-i\gamma\vv\cdot\kv}
  =\frac{-2\pi i\kv_v\cdot\zh\times\kv}
  {k^2(k^2-i\kv_v\cdot\kv)},
\label{theta_sk_moving}
\end{eqnarray}
where $\kv_v\equiv\gamma\vv/K$. This leads to the ``elastic'' energy
spectrum,
$|\nabla\theta(\kv)|^2=4\pi^2(k^2+k_v^2)/[(\kv_v\cdot\kv)^2+k^4]$,
that, on length scales beyond $\xi_v$ ($k\ll k_v$) is highly
anisotropic, akin to that of a smectic liquid crystal.  On shorter
length scales it reduces to that of an isotropic stationary
(undistorted) vortex\cite{CLbook}.

In real space the {\em steady}-state distortion for a $2\pi$-vortex
moving along the x-axis, in the vortex frame is given by
\begin{eqnarray}
\theta'_s(\rv)&\approx&-\int_{0}^\infty \frac{d q}{q}
\frac{e^{-q |\hat x|}\sin q \hat y}{q + 1}
\label{theta_sr}\\
&&-2\Theta(-\hat x)\int_{0}^\infty \frac{d q}{q}
\frac{e^{-q|\hat x|}\sin q \hat y}{q^2 - 1}
\left[1-e^{-q(q-1)|\hat x|}\right],\nonumber
\end{eqnarray}
where $\hat x = x/\xi_v$, $\hat y = y/\xi_v$.  Evaluating above
integrals numerically and adding the singular part of the vortex,
$\theta_v(\rv)=\varphi$, gives the real-space vector fields
illustrated in Figs.~\ref{fig:vortexantivortexComoving},
\ref{fig:vortexantivortexAnnihilate}.

A {\em transient}-state field of a vortex that has been moving for
time $t$ (particularly relevant for the annihilation problem) can also
be computed exactly and is given by
\begin{eqnarray}
  \theta'_s(\rv,t) &=&\int_\kv\vv\cdot\grad\theta_v(\kv)
  \frac{1-e^{-\frac{K}{\gamma}(k^2-i\kv_0\cdot\kv)t}}
  {k^2-i\kv_0\cdot\kv}e^{i\kv\cdot\rv}.\ \ \ \ 
\label{theta_r_transient}
\end{eqnarray}
Its key generic features are controlled by three length scales
$\xi_v$, $\xi_\perp\equiv\sqrt{K t/\gamma}$, $\xi_\parallel=v t$. At
time $t > t_*\equiv\xi_v/v$, such that $\xi_v\ll\xi_\perp\ll
\xi_\parallel$, one can see from the solution \rf{theta_r_transient}
that on scales shorter than an anisotropic domain
$\xi_\parallel\times\xi_\perp$, the solution reduces to the
``comet-tail'' steady-state one, \rf{theta_sr}
(Figs.~\ref{fig:vortexantivortexComoving},
\ref{fig:vortexantivortexAnnihilate}). On longer scales the vortex
distortion reduces to $\theta_s'(\rv,t)\approx\vv
t\cdot\grad\theta_v(\rv')$, which when combined with the singular part
gives
\begin{eqnarray}
  \theta'(\rv',t)\approx\theta_v(\rv') 
  + \vv t\cdot\grad\theta_v(\rv')\approx\theta_v(\rv'+\vv_0
  t)=\varphi.\ \ \ \ 
\label{undistorted_v}
\end{eqnarray}
Thus on scales outside of the $v t\times\sqrt{K t/\gamma}$ domain the
vortex field reduces to that of an undistorted stationary vortex
$\theta(\rv,t) = \varphi$ at its initial, $t=0$ position (see
Fig.\ref{fig:vortex_transient}). This is a diffusive vortex analog of
a ``causal horizon'' beyond which the distortion associated with a
moving vortex had not had sufficient time to propagate out.  Other
results (e.g., a vanishing vortex mobility, vortex energy and
interaction between moving vortices) follow directly from the above
moving vortex solution.

{\it Vortex mobility:} In the steady-state the power input by the
external force $F$ to drive the vortex at velocity $v$ is balanced by
the rotational power dissipated, $P_{rot}=\int_\rv(\partial_t\theta)
(K\nabla^2\theta)=\int_\rv\gamma(\partial_t\theta)^2 =\gamma
v^2\int_\rv(\partial_x\theta)^2$, gives the vortex drag coefficient,
$\gamma_v\equiv\mu^{-1}$ (inverse
mobility)\cite{JProst83,LRT86,RyskinKremenetskyPRL91}:
\begin{eqnarray}
\gamma_v&=&\gamma\int_{0}^{a^{-1}} dk k \int_0^{2\pi}
d\theta\frac{\sin^2\theta}
{k_v^2\cos^2\theta + k^2},\\
&=&\pi\gamma\sinh^{-1}\left(\frac{1}{k_v a}\right)
\approx\pi\gamma\ln(2\xi_v/a)\sim\gamma\ln v.\nonumber
\label{gamma_v}
\end{eqnarray}
Thus, at finite velocity, a previously noted divergence with system
size $L$ or vortex separation $r$\
\cite{PleinerPRA88,MuznyClarkPRL92,YurkeHusePRE93} is cutoff by the
velocity-length $\xi_v\sim 1/v$, thereby displaying a nonlinear
velocity-force characteristics i.e., an absence of linear response
down to a vanishing force.

{\it Vortex energy:} It is of interest to calculate the elastic energy
$E_v = \frac{K}{2}\int d^2r |\nabla\theta|^2$ stored in a moving
vortex.  In steady-state, using \rf{theta_sk_moving} I find:
\begin{eqnarray}
E_v
=\pi K\left(\sqrt{(L/\xi_v)^2+1} + \ln(\xi_v/a)
    -\sinh^{-1}(\xi_v/L)\right),\nonumber
\label{Evortex2}
\end{eqnarray}
that for vanishing velocity, $L\ll\xi_v$ reduces to $\ln L/a$ of a
stationary vortex, but for a rapidly moving vortex, $L\gg\xi_v$ gives
the energy \rf{Evortex}, that diverges {\em linearly} with $L$ and
with the standard logarithm cut off by the velocity-length $\xi_v$.
This later result is due to the confinement of the elastic distortion
(that in a stationary vortex is uniformly azimuthally distributed) to
a comet-tail wake of a moving vortex.

\begin{figure}[tbp]
\vspace{0.2cm}
\includegraphics[height=1.4in,width=3in]{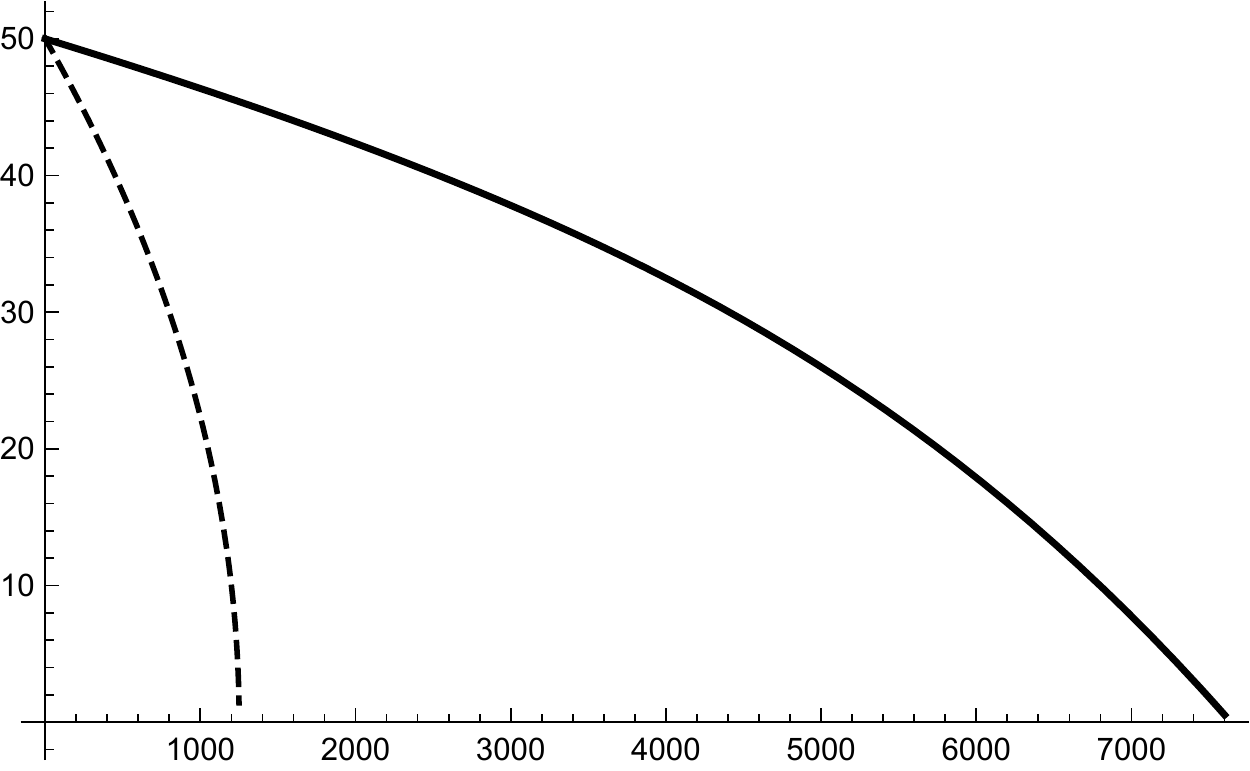}
\caption{Vortex-antivortex separation $r(t)$ as a function of time, a
  solution of \rf{EOMmodified} (solid), is significantly slowed down
  compared to the ``naive'' dynamics $\gamma_vdr/dt = -K/r$ (dashed).}
\label{fig:rt}
\end{figure}

{\it Vortex interaction:} To further characterize the nature of moving
vortices I study vortex-antivortex interaction, that strongly depends
on their velocities and orientation relative to the initial separation
vector, $\rv_\parallel=\rv_+-\rv_-$.

Motivated by the vortex-pair {\em annihilation} dynamics, I first
compute the energy $E_{\vv,-\vv}(r_\parallel)=\frac{K}{2}\int
d^2r|\nabla\theta_{\vv,-\vv}|^2$ of a vortex-antivortex pair moving
toward each other with velocity $\pm\vv=\pm v\hat{\bf r}$ along the separation
vector $\rv_\parallel$.
In steady-state the solution is given by
$\theta_{\vv,-\vv}(\rv,t) =\theta_{s}^+(\rv-\rv_+-\vv
t)+\theta_{v}^+(\rv-\rv_+-\vv t) +\theta_{s}^-(\rv-\rv_-+\vv
t)+\theta_{v}^-(\rv-\rv_-+\vv t)$, with singular (v) and smooth (s)
components for vortex (at $\rv_+$) and antivortex (at $\rv_-$),
respectively. The corresponding elastic energy
$E_{\vv,-\vv}(r_\parallel)=K\int\frac{d^2k}{k^2}\left[
1-e^{i\kv\cdot\rv_\parallel(t)}+(k^2k_v^2-(\kv_v\cdot\kv)^2)
\left(\frac{1}{k^4+(\kv_v\cdot\kv)^2}\right.\right.$   
$\left.\left.+\frac{e^{i\kv\cdot\rv_\parallel(t)}}{(k^2-i\kv_v\cdot\kv)^2}\right)\right],
$ is given by
\begin{widetext}
\begin{eqnarray}
E_{\vv,-\vv}(r_\parallel)
&\approx&
2\pi K\left[\frac{L}{\xi_v}
+\ln\frac{\xi_v}{a}-\sinh^{-1}\frac{\xi_v}{r_\parallel}\right]
\approx 2\pi K\left\{\begin{array}{ll}
L/\xi_v + \ln(\xi_v/a)-\xi_v/r_\parallel,
&a\ll\xi_v\ll r_\parallel,\\
L/\xi_v + \ln(r_\parallel/a),&a\ll r_\parallel\ll\xi_v,\\
\end{array}\right.
\label{Eparallel_vv}
\end{eqnarray}
\end{widetext}
where $\rv_\parallel(t) = \rv_+-\rv_- - 2\vv t$ and above I evaluated
the asymptotic $r_\parallel$ dependence using an approximate hard
cutoff $\xi_v/r_\parallel$ on low $q$.
Even for coinciding vortex-antivortex positions a linear in system
size contribution $L/\xi_v$ remains due to elastic energy associated
with the comet tail of each moving vortex (see
Fig.\rf{fig:vortexantivortexAnnihilate}). Subtracting this constant
self-energy piece I obtain the vortex-antivortex interaction,
$U^\parallel_{\vv,-\vv}(r_\parallel)$ advertised in \rf{Uparallel-vv},
that is qualitatively weaker and shorter range, falling off as
$1/r_\parallel$ at large separations, $r_\parallel\gg\xi_v$.

Before moving on, I stress that a full vortex annihilation problem is
far richer, requiring analysis of a full transient dynamics as
vortices accelerate from rest, with their velocity-length $\xi_v(t)$
evolving nontrivially and tails limited by the ``causal horizon'',
growing with $t$ from below to beyond their separation,
$r_\parallel(t)$. Consequently, the nature of the interaction
$U^\parallel_{\vv,-\vv}(r_\parallel,\xi_v)$ is nontrivially velocity
dependent. I analyze the associated dynamics of $\rv(t)$ below.

Another contrasting geometry of interest is that of a
vortex-antivortex pair co-moving (see
Fig.\ref{fig:vortexantivortexComoving}) with velocity $\vv$
perpendicular their separation vector $\rv_\perp=\rv_+-\rv_-$. In
steady-state, the solution $\theta_{\vv,\vv}(\rv,t)
=\theta_{s}^+(\rv-\rv_+-\vv t)+\theta_{v}^+(\rv-\rv_+-\vv t)
+\theta_{s}^-(\rv-\rv_--\vv t)+\theta_{v}^-(\rv-\rv_--\vv t)$ leads to
the elastic energy $E_{\vv,\vv}(r_\perp)=\frac{K}{2}\int
d^2r|\nabla\theta_{\vv,\vv}|^2=K\int
d^2k\frac{k^2+k_v^2}{k^4+(\kv_v\cdot\kv)^2}
\left(1-e^{i\kv\cdot\rv_\perp}\right)$ given by
\begin{widetext}
\begin{eqnarray}
E_{\vv,\vv}(r_\perp)&\approx&2\pi K\left[\ln\frac{\xi_v}{a}-\sinh^{-1}\frac{\xi_v}{r_\perp}
+ \sqrt{r_\perp^2/\xi_v^2+1}\right]
\approx 2\pi K\left\{\begin{array}{ll}
r_\perp/\xi_v + \ln(\xi_v/a),&a\ll\xi_v\ll r_\perp,\\
\ln(r_\perp/a),&a\ll r_\perp\ll\xi_v,\\
\end{array}\right.
\label{Eperp_vv}
\end{eqnarray}
\end{widetext}
evaluated in the same hard cutoff approximation as in
\rf{Eparallel_vv}, and giving $U^\perp_{\vv,\vv}(r_\perp)$ advertised
in \rf{Uperpvv}. This is a striking result as it predicts for $r_\perp
> \xi_v$ a linear confinement of a moving vortex-antivortex pair,
replacing logarithmic potential for a stationary pair. As is clear
from Fig.\rf{fig:vortexantivortexComoving} this elastic energy is
associated with the $r_\perp$ length of the non-overlapping parts of
the ``comet'' tails, the rest, beyond $r_\perp$ parts canceling
between co-moving vortex and antivortex.

{\it Vortex-antivortex annihilation dynamics}, approximately described
(neglecting\cite{neglectEffects} transients in \rf{theta_r_transient})
by $\gamma_v dr/dt = -\frac{\partial U_{-\vv,\vv}(r,v)}{\partial
  r}=-\frac{2\pi K}{r}\frac{1}{\sqrt{r^2/\xi_v^2+1}},$
\begin{eqnarray}
\dot\rh \ln(|\dot \rh|/2)
  &=& \frac{1}{\rh}\frac{1}{\sqrt{\rh^2\dot \rh^2+1}}
\label{EOMmodified}
\end{eqnarray}
is significantly enriched\cite{neglectEffects} by the
velocity-dependent mobility \rf{mu_result} and interaction
\rf{Uparallel-vv}, as compared to the naive dynamics $\gamma dr/dt =
-K/r$, that predicts a vortex separation $r(t) = \sqrt{r_0^2 -
  (2K/\gamma)t}$, initially separated by $r_0$, annihilating in time
$t_0 = r_0^2\gamma/(2K)$.\cite{ClarkExp}. Above $\rh$ and $\hat t$ are
respectively measured in the microscopic units of $a$ and $t_a =
a^2\gamma/(2K)$. Equation \rf{EOMmodified} predicts in units of
$v_a=a/t_0$ that $\rh \vh
=\frac{1}{\sqrt{2}}\left(\sqrt{1+4/\ln^2(\vh/2)}-1\right)^{1/2}$
(rather than $\rh \vh = const.$ of the naive dynamics) and can be
solved numerically, with the result illustrated in Fig.\rf{fig:rt}.
It shows a significant modification and slowing of the dynamics by the
effects studied here.



Beyond the transient time $\xi_v/v$, the enriched dynamics is expected
only in the regime of large separation and high velocity $r v \gg a
v_a = K/\gamma$, corresponding to $r \gg \xi_v$.  Using
$K/\gamma\approx 10^{-5}$ cm$^2$/sec and $v = 1 \mu$m/sec I estimate
$\xi_v\approx 1$mm and $v_a\approx 1$mm/sec for $a\approx 1\mu$m, a
limited regime of current experiment's\cite{ClarkExp}
applicability. Also, above prediction for the product $r v$ decreasing
with $r$ is inconsistent with measurements\cite{ClarkExp}. Thus, I
conclude that in current vortex annihilation experiments, the high
velocity effects studied here are not sufficient to account for the
observed anomalies\cite{ClarkExp} and other
effects\cite{neglectEffects} may need to be considered. Further
systematic experiments on moving vortices would be highly desirable
to sort out various contributions.

I also leave the extension of the present London limit analysis to a
superfluid, beyond a linearized xy-model treatment\cite{Arovas},
incorporating the full Galilean invariance\cite{Aleiner} for a future
study.

I thank Noel Clark for sharing his vortex annihilation data prior to
publication, and acknowledge him and Sandy Fetter, Bert Halperin, Dan
Arovas, and John Toner for stimulating discussions. This research was
supported by the NSF through DMR-1001240, MRSEC DMR-0820579, and by
the Simons Investigator award from the Simons Foundation.

\end{document}